\documentclass[12pt]{article}
\usepackage{fullpage}
\usepackage{graphicx}
\newcommand{\be}{\begin{equation}}
\newcommand{\ee}{\end{equation}}
\newcommand{\bphi}{\mbox{\boldmath $\phi$}}

\newcommand{\wh}{\widehat}

\newcommand{\cd}{\partial}

\newcommand{\ra}{\rightarrow}
\newcommand{\eps}{\mbox{\boldmath $\epsilon$}}
\newcommand{\C}{\bb{C}}
\newcommand{\ip}[1]{\langle#1\rangle}
\def \d{\mathrm{d}}
\def \tr{\mathrm{tr\,}}
\def \vol{\mathrm{vol}}

\def\v {{\bf v}}
\def\X {{\bf X}}

\def\bx {{\bf x}}
\def\bn {{\bf n}}

\def\by {{\bf a}}
\def\bq {{\bf q}}
\font\mybb=msbm10 at 11pt

\def\bb#1{\hbox{\mybb#1}}

\def\R {\bb{R}}

\def\Z {\bb{Z}}

\newcommand{\news}{\setcounter{equation}{0}\quad}
\def\ben{\begin{equation}}
\def\een{\end{equation}}
\def\beq{\begin{equation}}
\def\eeq{\end{equation}}
\def\bea{\begin{eqnarray}}
\def\eea{\end{eqnarray}}
\begin{document}
\title{
\begin{flushright}\ \vskip -2cm {\small{\em DCPT-11/23}}\end{flushright}
\vskip 2cm Broken Baby Skyrmions}
\author{Juha J\"aykk\"a$^\star$, Martin Speight$^\star$
and Paul Sutcliffe$^\dagger$\\[10pt]
{\em \normalsize $^{\star}$School of Mathematics,
University of Leeds, Leeds LS2 9JT, U.K.}\\
{\em \normalsize $^\dagger$Department of Mathematical Sciences,
Durham University, Durham DH1 3LE, U.K.}\\[10pt]
{\normalsize Email: \quad  
juhaj@iki.fi, \ speight@maths.leeds.ac.uk, \ p.m.sutcliffe@durham.ac.uk}
}
\date{September 2011}
\maketitle
\begin{abstract}
The baby Skyrme model is a (2+1)-dimensional analogue of the
Skyrme model, in which baryons are described by topological solitons.
In this paper we introduce a version of the baby Skyrme model in
which the global $O(3)$ symmetry is broken to the dihedral group
$D_N.$ It is found that the single soliton in this theory is 
composed of $N$ partons, that are topologically confined. 
The case $N=3$ is studied in some
detail and multi-soliton solutions are computed. 
These soliton solutions are related to  polyiamonds, which
are plane figures composed of equilateral triangles 
joined by common edges.
It is shown that the solitons may be
viewed as pieces of a doubly periodic soliton lattice.
An alternative model with $D_3$ symmetry is also introduced,
which has an exact explicit soliton lattice solution.
Soliton solutions are computed and compared in
the two $D_3$ theories.
Some comments are made regarding
the extension of these ideas to the Skyrme model.
\end{abstract}

\newpage
\section{Introduction}\news
The Skyrme model \cite{Sk} is a nonlinear field theory in which
baryons are described by topological solitons, called Skyrmions. 
The model has been obtained from quantum chromodynamics (QCD)
as a low energy effective field theory in the limit in which
the number of colours, $N,$ is large \cite{Wi}. 
More recently, the Skyrme model has been derived from
string theory, in the context of holographic QCD, again
in the large $N$ limit \cite{SS}. 

The number of colours, $N,$ appears in the Skyrme model
only as a coefficient of the Wess-Zumino term.
This plays an important role in the quantization of
Skyrmions, but at the classical level the
soliton solutions are blind to the value of $N;$
as the Wess-Zumino term does not contribute to the classical
energy. An interesting issue is whether it is possible 
to incorporate an effective small value of $N$ at the 
level of the classical soliton solution.

The baby Skyrme model \cite{PSZ} is a (2+1)-dimensional analogue
of the Skyrme model, that has proved to be a useful testing
ground for the study of Skyrmions. Soliton solutions of the
baby Skyrme and related models are also of interest in their
own right, within the context of condensed matter physics
\cite{SKKR,Yu}, where direct experimental observations can
be made.

In this paper we introduce a version of the baby Skyrme model in
which the global $O(3)$ symmetry is broken to the dihedral group
$D_N.$ It is shown that this reproduces some key features
expected of Skyrmions in a toy model associated with a small 
number of colours.
In particular, it is found that the single soliton in this theory is 
composed of $N$ partons, that are topologically confined. 
The case $N=3$ is studied in some
detail and multi-soliton solutions are computed.
The model admits a variety of stable multi-solitons that take 
the form of polyiamonds; which
are plane figures composed of equilateral triangles 
joined by common edges.
Doubly periodic soliton lattices are also computed
and it is shown that the polyiamond solitons 
may be viewed as pieces of a soliton lattice. 

An alternative model with $D_3$ symmetry is also
introduced and studied. This model has the property
that an exact soliton lattice solution 
is given explicitly in terms of a Weierstrass
elliptic function. Soliton solutions are computed and 
compared in the two $D_3$ theories.

\section{The broken baby Skyrme model}\news
The field of the baby Skyrme model is a three-component unit
vector ${\bphi}=(\phi_1,\phi_2,\phi_3).$
In this paper we are concerned with static solitons, hence the
theory may be defined by its static energy, which takes the form
\be
E=\int \bigg(\frac{1}{2}\partial_i\bphi\cdot\partial_i\bphi
+\frac{\kappa^2}{4}(\partial_i\bphi\times\partial_j\bphi)\cdot
(\partial_i\bphi\times\partial_j\bphi)+V\bigg)\,d^2x,
\label{energy}
\ee
where $V(\bphi)$ is a potential and $\kappa$ is a constant.
In the standard baby Skyrme model \cite{PSZ} the potential is taken to
be 
\be
V=m^2(1-\phi_3),
\label{pot1}
\ee
which is the analogue of the conventional pion mass term
in the Skyrme model \cite{AN}. The constant $m$ gives the
mass of the fields $\phi_1$ and $\phi_2,$ associated with
elementary excitations around the unique vacuum $\bphi=(0,0,1).$

Finite energy requires that the field takes the vacuum value
at all points at spatial infinity, 
$\bphi(\infty)=(0,0,1).$
This compactification means that
topologically $\bphi$ is a map between two-spheres, with an
associated integer winding number $B\in\bb{Z}=\pi_2(S^2).$
This topological charge (or soliton number) is the analogue of the
baryon number in the Skyrme model and may be calculated as
\be
B=-\frac{1}{4\pi}\int \bphi\cdot(\partial_1\bphi\times\partial_2\bphi)\, d^2x.
\label{charge}
\ee
An application of Derrick's theorem \cite{De} reveals that the 
scale of a soliton in the baby Skyrme model is determined by the ratio
$\sqrt{\kappa/m}.$

The first two terms in the energy (\ref{energy}) are invariant 
under the global $O(3)$ symmetry, $\bphi\mapsto {\cal O}\bphi$ for
${\cal O}\in O(3),$ but this is broken by the potential 
(\ref{pot1}) to an $O(2)$ symmetry acting on the first two components 
$\phi_1,\phi_2.$ The single soliton takes advantage of this symmetry
and is axially symmetric \cite{PSZ}.

Other choices for the potential have been investigated 
\cite{LPZ2,We}, but in these examples there is an 
unbroken
$O(2)$ symmetry and the $B=1$ soliton is axially symmetric.
A novel situation was considered recently \cite{JS}
using the easy plane potential $V=\frac{1}{2}m^2\phi_1^2.$ 
Again this leaves an unbroken $O(2)$ symmetry, but in this
case the choice of a vacuum value at spatial infinity,
for example $\bphi(\infty)=(0,0,1),$ distinguishes a point
on the orbit of the unbroken symmetry and
further breaks the symmetry to the dihedral group $D_2.$ 
In this case the single soliton is not axially 
symmetric, but turns out to be composed of two constituents. 
The work on easy plane baby Skyrmions provided
some inspiration for the model proposed in the current paper, 
but it should be stressed that the two theories are quite different,
as described shortly.

The only previous work we are aware of in which the 
potential has only a discrete symmetry is the work
of Ward \cite{Wa}, in which the symmetry is broken
to the dihedral group $D_2$
by the choice 
$V=\frac{1}{2}m^2(1-\phi_3^2)(1-\phi_1^2).$
In this theory the single soliton is also composed
of two constituents.
 We shall discuss this
theory in more detail in section \ref{sec-alt}, where
we introduce a generalization to a model with 
$D_3$ symmetry and compare the results with those
obtained for the theory of main concern in this paper.

The theory introduced in the present paper
has the potential
\be
V=m^2\,|1-(\phi_1+i\phi_2)^N|^2\,(1-\phi_3),
\label{pot2}
\ee
where $N\ge 2$ is an integer parameter of the model.
 We shall refer to the theory
with this potential as the broken baby Skyrme model
with $N$ colours, to use a suggestive notation. 
The global $O(3)$ symmetry is broken by this potential 
to the dihedral group
$D_N,$ generated by
the rotation $(\phi_1+i\phi_2)\mapsto (\phi_1+i\phi_2)e^{i2\pi/N}$ 
and the reflection $(\phi_1,\phi_2,\phi_3)\mapsto(\phi_1,-\phi_2,\phi_3).$
 In the context of
symmetry groups in three dimensions, this pyramidal symmetry 
group is often denoted $C_{Nv},$ but as an abstract group
it is isomorphic to $D_N,$ which is a more convenient
notation for the later application to planar symmetries.

The potential (\ref{pot2}) has $N+1$ vacua on the two-sphere.
The vacuum at the north pole $\bphi=(0,0,1),$ will be the chosen
vacuum at spatial infinity, with the remaining $N$ vacua lying
on the equatorial circle $\phi_3=0$ at the $N$th roots of unity.
Note that this choice of boundary condition does not break any
further symmetries, in contrast to the situation for the
easy plane potential. 
Another important difference between the two
theories concerns the masses of the elementary excitations of
the fields $\phi_1$ and $\phi_2.$ To quadratic order in 
$\phi_1$ and $\phi_2,$ the broken potential (\ref{pot2}) agrees with
the standard baby Skyrme potential (\ref{pot1}), hence both
fields $\phi_1$ and $\phi_2$ have mass $m.$ However, for the
easy plane potential clearly only the $\phi_1$ field has mass $m$
and the $\phi_2$ field is massless. Hopefully, this brief discussion
has served to highlight some of the important differences between the two
models, making it clear that different phenomena should be expected, 
even in the case $N=2$ where both theories have the same unbroken 
symmetry group $D_2.$ 

Ward's potential \cite{Wa} with $D_2$ symmetry has
more in common with the two colour broken baby Skyrme model,
as both fields $\phi_1$ and $\phi_2$ have the same mass.
However, there is an additional inversion symmetry
$\bphi\mapsto-\bphi,$ with an associated extra 
vacuum  $\bphi=(0,0,-1),$ which leads to some
qualitative differences in the soliton solutions.

\section{Solitons and polyiamonds}\news\label{sec-sol}
Whether in two or three dimensions, a symmetry of a soliton in a 
Skyrme model refers to an equivariance in which a spatial rotation
or reflection can be compensated by the action of the global symmetry
of the theory. Clearly the maximal symmetry possible in the broken
baby Skyrme model with $N$ colours is dihedral symmetry $D_N.$
Even the $B=1$ single soliton cannot be axially symmetric, though
the expectation is that it has the maximal symmetry $D_N.$
Numerical solutions presented in this section confirm this
expectation and reveal that the single soliton is composed of 
$N$ constituents, which we refer to as partons, given that in the
Skyrme model the single soliton describes the proton. 

Each parton carries baryon number $B=1/N,$ associated with a winding
that covers this fraction of the target two-sphere. Partons are
topologically confined, since finite energy requires that the 
total baryon number is integer-valued, hence an equal number of partons
of each colour. Here the term colour may be used to enumerate the
$N$ segments of the target two-sphere obtained by drawing great
semi-circles from the north pole to the south pole that pass through
each of the $N$ equatorial vacua. Of course a parton and an anti-parton
(associated with a winding of the same segment with the opposite
orientation) is an allowed combination, corresponding to elementary
excitations of the $\phi_1$ and $\phi_2$ fields with zero baryon number,
being the baby Skyrme analogue of pions. 

From now on we shall concentrate on the 
most physically relevant case of $N=3$ colours, 
regarding the theory as a lower-dimensional analogue of 
the Skyrme model. 
The generic values $\kappa=m=1$ are taken for the parameters
of the theory.

To numerically construct soliton solutions 
an energy minimizing gradient flow algorithm is applied to
the energy (\ref{energy}).
Spatial derivatives are approximated using fourth-order
accurate finite difference approximations with a 
lattice spacing $\Delta x=0.05$ and $251^2$ grid points.
At the boundary of the grid the field is fixed to the
vacuum value $\bphi=(0,0,1).$

A field with topological charge $B$ is given by
\be
\bphi=(\sin f\cos(B\theta),\sin f\sin(B\theta), \cos f),
\label{ic}
\ee
where $r$ and $\theta$ are polar coordinates in the plane
with $f(r)$ any monotonically decreasing radial profile function
such that $f(0)=\pi$ and $f$ vanishes at the boundary of the grid.

The field (\ref{ic}) has dihedral symmetry $D_{3B},$ with the 
spatial rotation
$\theta\mapsto\theta+2\pi/(3B)$ being compensated by 
the global transformation 
$(\phi_1+i\phi_2)\mapsto (\phi_1+i\phi_2)e^{-i2\pi/3},$ and
the spatial reflection $\theta\mapsto -\theta$ compensated by
the global reflection 
$(\phi_1,\phi_2,\phi_3)\mapsto(\phi_1,-\phi_2,\phi_3).$
Note that either a spatial reflection or a global reflection
alone changes the sign of the topological charge, but the
combination of the two leaves it unchanged.

\begin{figure}[ht]\begin{center}
\includegraphics[width=17cm]{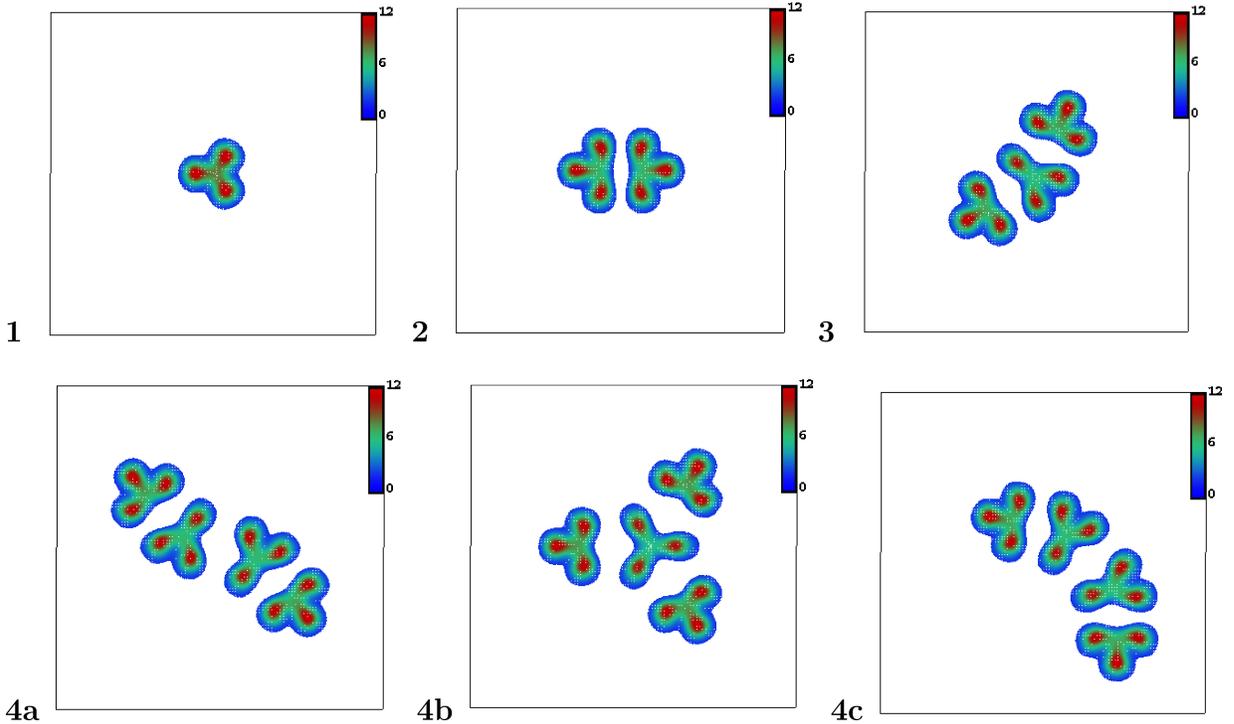}
\caption{Energy density contour plots for solitons with
charge $B.$ The top row displays stable solitons
for $B=1,2,3$ and the bottom row shows three different
stable solitons with $B=4.$}
\label{fig-all6}\end{center}\end{figure}

Using the field (\ref{ic}) as an initial condition in the
numerical energy minimization code yields a charge $B$ solution
with $D_{3B}$ symmetry. The energy density of the $B=1$ soliton 
is displayed as a contour plot in Figure~\ref{fig-all6}.1, which
clearly shows the three constituent partons. A plot of the 
topological charge density (the integrand in (\ref{charge})) has
a similar structure. The energy of this $D_3$ symmetric $B=1$
soliton is $E=34.79,$ and is listed in Table~\ref{tab-energies}.
For all the numerical results presented in this paper, the 
topological charge computed using the lattice version of
(\ref{charge}) is integer-valued to five significant figures,
which provides an indication of the accuracy expected in the
numerical computations.

\begin{figure}[ht]\begin{center}
\includegraphics[width=16cm]{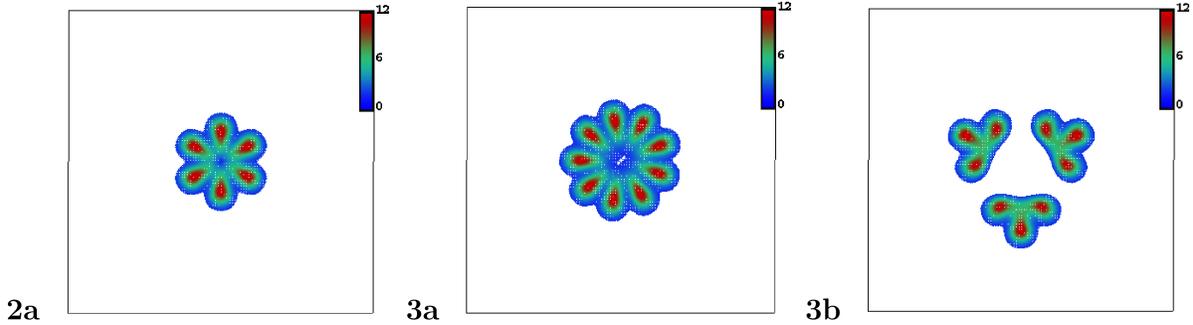}
\caption{Energy density contour plots for solitons with
charge $B.$ \hskip 3cm \break
2a) $B=2$ hexagon; \ 3a) $B=3$ nonagon; \ 3b) $B=3$ triquetra.
}
\label{fig-nonpoly}\end{center}\end{figure}

\begin{table}[ht]
\centering
\begin{tabular}{|c|c|c|l|}
\hline
$B$ & $E/B$  &  $G$ &  Figure\\ \hline
1 & 34.79 & $D_3$ & \ref{fig-all6}.1\\
2 & 33.04 & $D_6$ & \ref{fig-nonpoly}.2a\\
2 & 33.06 & $D_2$ & \ref{fig-all6}.2\\
3 & 32.83 & $D_1$ & \ref{fig-all6}.3 \\
3 & 33.00 & $D_3$ & \ref{fig-nonpoly}.3b \\
3 & 33.68 & $D_9$ & \ref{fig-nonpoly}.3a \\
4 & 32.68 & $C_2$ & \ref{fig-all6}.4a \\
4 & 32.70 & $D_1$ & \ref{fig-all6}.4c \\
4 & 32.96 & $D_3$ & \ref{fig-all6}.4b \\
\hline
\end{tabular}
\caption{The energy per soliton $E/B$ and symmetry group $G$ of solitons
with topological charge $B\le 4.$
}
 \label{tab-energies}
\end{table}

\begin{figure}[ht]\begin{center}
\includegraphics[width=10cm]{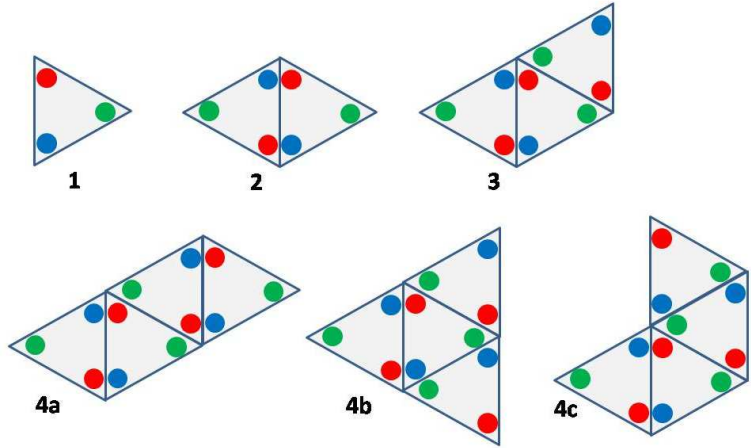}
\caption{
The six possible polyiamonds constructed from at most
four triangles. }
\label{fig-polyiamonds}\end{center}\end{figure}

The $D_{3B}$ symmetric solution is composed of 
$3B$ partons located at the vertices of a regular $3B$-gon.
The solutions with $B=2$ and $B=3$ are displayed in 
Figures~\ref{fig-nonpoly}.2a and ~\ref{fig-nonpoly}.3a 
respectively.
For $B>2$ such solutions are unstable to perturbations that break the
dihedral symmetry. The $B=2$ hexagonal solution is stable 
with an energy per soliton of $E/B=33.04.$ 
A sufficiently large symmetry breaking perturbation can convert this
solution into the additional stable 
$D_2$ symmetric soliton displayed in Figure~\ref{fig-all6}.2.
The energy per soliton of this solution is
$E/B=33.06,$ so the two different $B=2$ solutions have 
energies that differ by an amount comparable to our 
expected numerical accuracy, and we are unable to make a 
confident statement about which has the lower energy.
It is clear from Figure~\ref{fig-all6}.2
 that this $B=2$ soliton
is constructed from two $B=1$ solitons, with a relative 
spatial rotation of $180^\circ.$
 
As mentioned earlier, the asymptotic fields of a soliton in
the broken baby Skyrme model have the same form as those in
the standard baby Skyrme model; hence the results on asymptotic
interactions derived in \cite{PSZ} can be transfered to the
current theory (see Appendix A for a discussion
of asymptotic forces for a general potential).
 In particular, the leading order result 
shows that two single solitons are maximally attractive if one
is rotated relative to the other through an angle of $180^\circ.$
As the single soliton in the broken baby Skyrme model is not 
axially symmetric, then beyond leading order there will be 
a contribution that differentiates between the relative orientation
of the two triangles of partons. 
The result presented in Figure~\ref{fig-all6}.2 displays the optimal
orientation between the two triangles, which can be confirmed by 
using an initial condition consisting of two well-separated 
single solitons with a generic initial orientation.

It is instructive to represent the single soliton, 
as in Figure~\ref{fig-polyiamonds}.1,
 as a triangle with three coloured dots to denote the three
peaks in the energy density associated with the three segments
of the target two-sphere. With this representation the 
$B=2$ soliton in Figure~\ref{fig-all6}.2 corresponds to the diamond in 
Figure~\ref{fig-polyiamonds}.2, in which the two triangles
share a common edge with adjacent dots being different colours.

This triangle representation suggests that multi-solitons will
correspond to polyiamonds. 
A polyiamond \cite{OB} is a plane figure composed of
identical equilateral triangles joined by common edges, so that no
two triangles overlap. For three triangles there is a 
unique triamond, shown in Figure~\ref{fig-polyiamonds}.3.
Given a polyiamond, and a colour assignment to any one of the
triangles, there is a unique colouring of the triangles
satisfying the rule of different adjacent colours.

The initial condition (\ref{ic}) with $B=3$ yields the
unstable $D_9$ symmetric nonagon displayed in
Figure~\ref{fig-nonpoly}.3a  with an energy per soliton of 
$E/B=33.68.$
A symmetry breaking perturbation leads to the $B=3$ soliton
displayed in Figure~\ref{fig-all6}.3 with $E/B=32.83.$
This confirms the predicted triamond form of 
Figure~\ref{fig-polyiamonds}.3, which has only a 
reflection symmetry, $D_1,$ and no rotational symmetry. 
This $B=3$ soliton can also be obtained from an initial 
condition consisting of three well-separated single solitons
with appropriate positions and orientations.
A stable $B=3$ local energy minimum has also been computed that is not
of the polyiamond form. 
It has an energy per soliton of $E/B=33.00$ and is
presented in Figure~\ref{fig-nonpoly}.3b. This triquetra solution 
has $D_3$ symmetry and is formed from three triangles
that share vertices but not edges. 

With four triangles there are the three tetriamonds shown in 
Figures \ref{fig-polyiamonds}.4a, \ref{fig-polyiamonds}.4b and 
\ref{fig-polyiamonds}.4c. 
Suitable initial conditions, using four well-separated single
solitons with appropriate positions and orientations, 
leads to $B=4$ solitons associated with each of these
tetriamonds. Energy density plots for all three solitons
are displayed in the bottom row of 
Figure~\ref{fig-all6} and the associated 
energies are listed in Table~\ref{tab-energies}.
All three solutions appear to be stable and their energies
are very close to each other. This can be understood from the
fact that all pairs of triangles that share a common edge are
in a maximally attractive orientation. Furthermore, to 
transform from one configuration to another requires the 
breaking of an attractive bond followed by a change of orientation
and a repositioning of a triangle. 

The number of polyiamonds grows rapidly with the number of
triangles, and the expectation is that there will be a soliton
associated with each of these. There are five pentiamonds
(including the first asymmetric configuration with neither a
rotation nor a reflection symmetry) and twelve hexiamonds.
Already for $B=9$ there are 160 polyiamonds, so it is likely 
to be a computationally intensive task to numerically compute
all the solitons for values of $B$ larger than those considered
in this paper. However, it could be a worthwhile exercise that
might lead to an interesting energy function on the space of 
polyiamonds. Based on the result for $B=4,$ it could be 
that the linear arrangement of triangles is minimal for all
values of $B.$ This would have some similarities with the standard
baby Skyrme model, where soliton chains are found to be the 
minimal energy configurations \cite{Fo}.

Note that the $B=3$ triquetra solution in Figure~\ref{fig-nonpoly}.3b is
related to the $B=4$ tetriamond solution in Figure~\ref{fig-all6}.4b
by removing the central triangle. 
A local energy minimum with $B=4$ has been computed that is not
of the polyiamond form but is related to the triquetra solution 
by the addition of a triangle (in the polyiamond manner) on
an outside edge of the triquetra, rather than in the middle.
It is expected that a variety of similar local energy minima exist for
all $B\ge 4,$ based on extending the triquetra solution in this manner.

\section{Soliton lattices}\label{sollat}
The polyiamond solutions 
in Figure~\ref{fig-all6}
suggest the existence of a doubly
periodic triangular lattice, associated with a tiling of the
plane by equalateral triangles. To study a lattice with
triangular symmetry it is sufficient to restrict to the
fundamental torus $\bb{T}^2$ with a $60^\circ$ angle 
in the $(x_1,x_2)$ plane.
It is useful to note that several identities may be derived
for such a lattice to be a critical point of the energy (\ref{energy})
(see Appendix B).

The first identity is a standard virial relation that follows from an 
application of Derrick's theorem \cite{De} and is a requirement
of criticality under a rescaling of the lattice
\be
\int_{\bb{T}^2} 
\bigg(
\frac{\kappa^2}{4}(\partial_i\bphi\times\partial_j\bphi)\cdot
(\partial_i\bphi\times\partial_j\bphi)-V\bigg)\,d^2x=0.
\label{identity1}
\ee 
The two remaining identities follow from variations of the lattice
associated with a stretch of one of the fundamental periods and
a variation of the angle of the fundamental torus from $60^\circ$
\be
\int_{\bb{T}^2} 
\bigg(
\partial_1\bphi\cdot\partial_1\bphi-
\partial_2\bphi\cdot\partial_2\bphi
\bigg)
\,d^2x=0,
\label{identity2}
\ee 
and
\be
\int_{\bb{T}^2} 
%\bigg(
\partial_1\bphi\cdot\partial_2\bphi
%\bigg)
\ d^2x=0.
\label{identity3}
\ee 
For computational purposes it is convenient to work with a
rectangular torus of the form
$(x_1,x_2)\in[0,L]\times[0,\sqrt{3}L],$ 
containing two copies of the fundamental torus.
Numerical simulations can then be performed in the rectangular
torus with periodic boundary conditions in both the $x_1$ and
$x_2$ directions. An initial value is chosen for the length $L$ 
of the torus and the energy minimized using the numerical methods
described in the previous section. The virial relation (\ref{identity1})
is then evaluated and used to predict an improved estimate for
the torus length $L.$ This procedure is iterated to 
convergence and finally the two remaining identities (\ref{identity2})
and (\ref{identity3}) are checked. For the results presented in
this paper the rectangular lattice contains 
$52\times 90$ grid points 
and all three identities are satisfied to an accuracy of
better than $1\%.$  

\begin{figure}[ht]\begin{center}
\includegraphics[width=7cm]{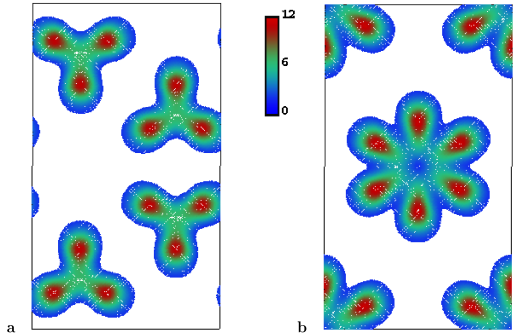}
\caption{Energy density contour plots for two different
soliton lattices. On the left is a triangular lattice
of single solitons and on the right is a hexagonal lattice of double solitons.
The single soliton lattice has a slightly lower energy per soliton.}
\label{fig-lat12}\end{center}\end{figure}

The fundamental torus of a triangular tiling contains two
triangles, hence the rectangular torus contains a field
with topological charge $B=4.$ An initial condition is provided
by setting $B=4$ in the ansatz (\ref{ic}) and using a radial profile
function with a compact support inside the rectangular torus.
A periodic perturbation is applied to this initial condition, to
break any reflection symmetries. The result of the numerical
minimization is displayed in the left-hand side of Figure~\ref{fig-lat12}.
The minimizing torus length is $L=5.38,$ with 
an energy per soliton for this lattice equal to $E/B=31.83.$ 
This value is consistent as a limit of the finite $B$  
polyiamond energies presented in Table~\ref{tab-energies}.

There is a second soliton lattice, based on the hexagonal $B=2$ soliton
of Figure~\ref{fig-nonpoly}.2a. This lattice is obtained if the initial
condition is not subjected to a perturbation that breaks the left-right
reflection symmetry.  
This lattice is displayed in the right-hand side of Figure~\ref{fig-lat12}.
In this case the minimizing torus length is slightly reduced at $L=5.10$ 
and the energy per soliton is a little greater at 
$E/B=32.43.$ The fact that the energy of the double soliton lattice
is larger than that of the single soliton lattice agrees with the
preferred polyiamond form of the minimal energy solitons with $B>2.$ 

All the polyiamond solutions in Figure~\ref{fig-all6} may be viewed
as finite $B$ pieces cut from the soliton lattice presented
in the left-hand side of Figure~\ref{fig-lat12}. This explains 
why different solutions with the same value of $B$ have very similar
energies. The difference is an edge effect associated with 
the way in which a finite $B$ piece is cut from the infinite lattice.
This supports the expectation that stable soliton solutions exist 
for each polyiamond and suggests that the energy differences 
between solutions will
decrease as $B$ increases.

\section{An alternative $D_3$ model}\news\label{sec-alt}
As mentioned earlier, 
in \cite{Wa} Ward introduced a potential with $D_2$ symmetry.
It has 
the novel feature that an exact explicit solution exists for
the soliton lattice, with the single soliton composed of two partons. 
The lattice has square symmetry and consists of half solitons.
In this section, we apply Ward's analysis to construct a potential
with $D_3$ symmetry, for which there is an exact explicit triangular lattice
solution. We then investigate the solitons of this alternative
$D_3$ model, pointing out some of the differences and similarities
with our earlier three colour theory.

Consider the energy (\ref{energy}) 
defined on a torus $\bb{T}^2,$
with an arbitrary potential
$V.$ 
Let $b$ denote the topological charge density, the integrand
in (\ref{charge}),
\be
b=-\frac{1}{4\pi}\bphi\cdot(\partial_1\bphi\times\partial_2\bphi).
\label{density}
\ee
The inequality
\be
(
4\pi\kappa b-\sqrt{2V}
)^2\ge 0
\label{square}
\ee
leads to the following lower bound for the energy on the torus
\be
E\ge 4\pi\bigg(
B+\kappa\int_{\bb{T}^2} b\sqrt{2V}\,d^2x
\bigg),
\label{bound0}
\ee
where, without loss of generality, we have restricted
to the situation where $B>0.$

Following \cite{Wa}, 
the inequality (\ref{bound0}) can be written 
as a Bogomolny bound
\be
E/B\ge 4\pi+\kappa\beta,
\label{bound1}
\ee
where $\beta$ is the constant
\be
\beta=\int_{S^2}\sqrt{2V}\, d^2S,
\label{beta}
\ee
with $d^2S$ the standard area element on the target two-sphere,
normalized to $4\pi.$

To investigate the bound further, it is useful to introduce complex
coordinates on both the domain and target by defining 
$z=x_1+ix_2$ and $W=(\phi_1+i\phi_2)/(1-\phi_3).$ 
A necessary condition to attain the bound (\ref{bound1})
is that $W$ is a meromorphic function of $z.$
With this assumption the topological charge density
(\ref{density}) becomes
\be
b=\frac{1}{\pi(1+|W|^2)^2}
\bigg|\frac{dW}{dz}\bigg|^2.
\label{density2}
\ee 
The bound is then attained if the inequality (\ref{square})
becomes an equality, which requires
\be
V=\frac{8\kappa^2}{(1+|W|^2)^4}\bigg|\frac{dW}{dz}\bigg|^4.
\label{defv1}
\ee
This defines a suitable potential if $dW/dz$ 
is an explicit function of $W.$
As the domain is a torus, the most elegant 
choice is to take $W$ to be proportional to a Weierstrass 
elliptic function. For simplicity, we set 
$W(z)=\wp(z),$ where the Weierstrass  
function is defined in terms of the invariants $g_2,g_3$ by
\be
\bigg(\frac{d\wp}{dz}\bigg)^2=4\wp^3-g_2\wp-g_3.
\label{defwp}
\ee
Substituting this choice into (\ref{defv1})
yields the family of potentials
\be
V=\frac{8\kappa^2|4W^3-g_2W-g_3|^2}{(1+|W|^2)^4}.
\label{defv2}
\ee
Ward studied the case $g_3=0,$ when
the potential (\ref{defv2}) has $D_2$ symmetry
($W\mapsto iW$),
with an exact lattice solution of half solitons and
square symmetry \cite{Wa}. In this paper we consider
the theory with $g_2=0,$ so that there is 
$D_3$ symmetry ($W\mapsto e^{i2\pi/3}W$).
Different values of $g_3$ are related by a scaling
symmetry, which we use to set $g_3=4.$
The potential of our alternative $D_3$ model is therefore
given by
\be
V=\frac{128\kappa^2|1-W^3|^2}{(1+|W|^2)^4}
=16\kappa^2(1-\phi_3)
(1+3\phi_3^2+3\phi_1\phi_2^2-\phi_1^3),
\label{defv3}
\ee 
where in the final expression the potential is 
written in terms of the field $\bphi.$

As in the previous sections, from now on we set 
$\kappa=1.$ 
The vacua of the potential (\ref{defv3}) 
are the same as those of the three colour broken baby Skyrme
model (\ref{pot2}).
Considering elementary excitations
around the vacuum $\bphi=(0,0,1),$ reveals that the
fields $\phi_1$ and $\phi_2$ both have mass $m=8.$
This is not an important difference, and is 
simply a consequence of our choice of 
scaling for the exact elliptic function solution,
but does mean that the scale of the solitons will be
smaller in the alternative $D_3$ model and the
energies higher. 

A significant difference between the two $D_3$ theories
concerns the relative values of the potential at
particular points of interest on the target two-sphere. 
These four points are $\bphi=(0,0,-1),$ defining the 
centre of the soliton, and the three points on the 
equator midway between the equatorial vacua, such
as $\bphi=(-1,0,0).$ In the broken baby Skyrme model
the potential at a midway point is greater than the
potential at the soliton centre, producing a split
of the single soliton energy density into three peaks.
However, for the alternative $D_3$ theory this situation
is reversed, so the energy density is not split but
is simply stretched into a triangular deformation.
In this sense the alternative model does not reveal
the parton constituents in the same way as the
broken baby Skyrme model. However, as we now describe,
there are a number of similarities between the solitons
of the two theories.

\begin{figure}[ht]\begin{center}
\includegraphics[width=3cm]{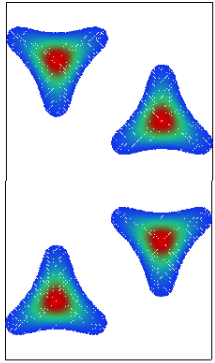}
\caption{An energy density contour plot of the 
lattice of single solitons given by an
 exact elliptic function solution of the 
alternative $D_3$ model.}
\label{fig-lat3}\end{center}\end{figure}

The energy density of the exact elliptic function
solution is displayed in Figure~\ref{fig-lat3}.
To aid comparison with the previous lattice solutions,
the energy density is plotted for a rectangular
torus containing two copies of the fundamental torus.
The length of the rectangular torus is equal to
the real period of the elliptic function, hence 
$L=\omega_1=\Gamma(\frac{1}{6})\Gamma(\frac{1}{3})/
(2\sqrt{3\pi})=2.428...$

A numerical evaluation of the integral (\ref{beta})
yields $\beta=89.30$ and hence from (\ref{bound1})
an energy per soliton of $E/B=101.86.$ 
Note the similarity between Figure~\ref{fig-lat3}
and the left-hand side of Figure~\ref{fig-lat12}.
Both are triangular lattices of single solitons,
with the main difference being that the energy density
of the single soliton is not split into three peaks in the
alternative $D_3$ theory, as discussed above. 
The similarity between the minimal energy lattices
in the two theories suggests that solitons in the
alternative $D_3$ theory may also be related to 
polyiamonds. To investigate this issue, the same
numerical code used in section \ref{sec-sol} is
applied to the problem, with the only change being
a reduction in the lattice spacing to $\Delta x=0.02,$
reflecting the smaller scale of the solitons. 

\begin{figure}[ht]\begin{center}
\includegraphics[width=16cm]{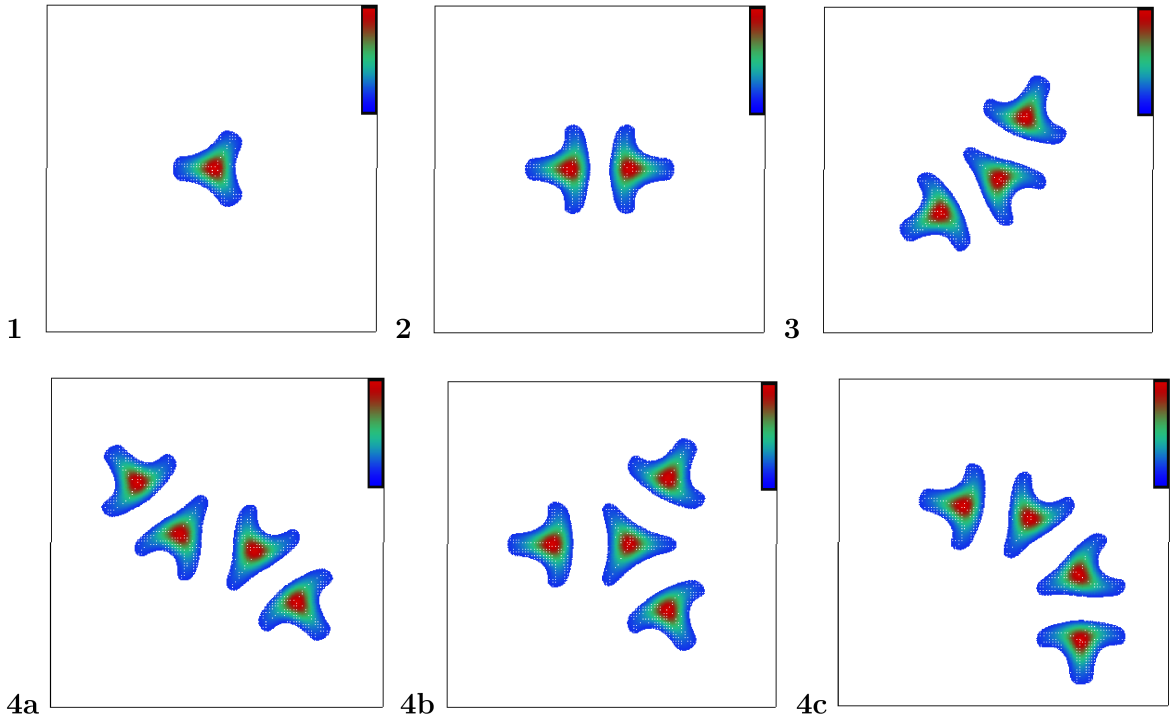}
\caption{Energy density contour plots for solitons 
of charge $B$ 
in the alternative $D_3$ model.
The top row displays stable solitons
for $B=1,2,3$ and the bottom row shows three different
stable solitons with $B=4.$}
\label{fig-eall6}\end{center}\end{figure}

\begin{table}[ht]
\centering
\begin{tabular}{|c|c|c|l|}
\hline
$B$ & $E/B$  &  $G$ &  Figure\\ \hline
1 & 108.99 & $D_3$ & \ref{fig-eall6}.1\\
2 & 105.80 & $D_2$ & \ref{fig-eall6}.2\\
3 & 104.99 & $D_1$ & \ref{fig-eall6}.3 \\
4 & 104.59 & $C_2$ & \ref{fig-eall6}.4a \\
4 & 104.62 & $D_1$ & \ref{fig-eall6}.4c \\
4 & 104.81 & $D_3$ & \ref{fig-eall6}.4b \\
\hline
\end{tabular}
\caption{The energy per soliton $E/B$ and symmetry group $G$ of solitons
with topological charge $B\le 4$ in the alternative $D_3$ model.
}
 \label{tab-eenergies}
\end{table}

Energy density contour plots are displayed in
Figure~\ref{fig-eall6} for some solitons with
$B\le 4.$ The similarity to Figure~\ref{fig-all6}
for the broken baby Skyrme model is obvious, as is
the polyiamond form of the solutions.
The symmetry and energy per soliton for each of these
solutions is presented in Table~\ref{tab-eenergies}
and the results are consistent with the limiting
lattice value of $E/B=101.86.$

One difference between the two theories concerns
the non-polyiamond solutions, such as the $B=2$
hexagon solution shown in Figure~\ref{fig-nonpoly}.2a.
In the alternative $D_3$ model a $B=2$ hexagon
appears to be unstable, with an energy per soliton
of $E/B=106.33,$ which is larger than that of the
diamond solution $E/B=105.80.$

%\newpage
\section{Conclusion}\news
A version of the baby Skyrme model has been introduced in which
the global $O(3)$ symmetry is broken by the potential to
a dihedral symmetry $D_N,$ with the result that the single
soliton is composed of $N$ topologically confined partons.
Multi-solitons have been computed in the $N=3$ theory 
and shown to be related to polyiamonds, with a form consistent
with cutting pieces from a doubly periodic soliton lattice.
It would be worth
extending the results in this paper to larger soliton numbers, to
verify that the polyiamonds correspondence continues.
It might also be amusing
to investigate the $N=4$ theory and determine whether the 
multi-solitons are related to polyominoes.

The obvious extension of the work in this paper
is to the (3+1)-dimensional Skyrme model. 
It is straightforward to construct an analogous symmetry 
breaking potential in the Skyrme model, 
though the physical consequences of isospin symmetry breaking
 must be carefully considered. 
The potential term in the Skyrme model does not play as
crucial a role as in the baby Skyrme model, so the 
influence of a symmetry breaking potential is not expected to
be as dramatic. However, even the inclusion of the 
traditional pion mass term does produce qualitative differences 
in the structure of multi-Skyrmions, for sufficiently large
baryon numbers \cite{BS-pm,BMS}, so some new features
should appear. 

\section*{Appendix A: long range inter-lump forces}

In this appendix we present an analysis of the long-range forces between
solitons in the baby Skyrme model with a general smooth potential 
$V:S^2\ra [0,\infty)$.
Let $\v\in V^{-1}(0)$ be the vacuum value of $\bphi$ 
(so $\lim_{r\ra\infty}\bphi=\v$). By definition,
this is a (possibly degenerate) minimum of $V$, so the Hessian of $V$ at $\v$ 
is well-defined, and has non-negative eigenvalues $\mu_1^2,\mu_2^2\geq 0$, 
which we choose to order so that $\mu_1^2\leq\mu_2^2$. 
(Recall that the Hessian of $V$ at $\v$ is the unique self-adjoint linear
map $J_\v:T_\v S^2\ra T_\v S^2$ such that
\beq
\frac{d^2 V(\bphi(t))}{dt^2}\bigg|_{t=0} \equiv \X\cdot J_\v \X,
\eeq
where $\bphi(t)$ is any curve in $S^2$ with $\bphi(0)=\v$, 
$\dot{\bphi}(0)=\X$.) The case $\mu_1^2=0$ was
treated in \cite{JS}, so we shall concentrate on the case where
 $\mu_1^2>0$. Let $\eps_1,\eps_2$ be
the corresponding unit eigenvectors, oriented so that $\eps_2=\v\times\eps_1$. 
At large $r$ one
expects the fields of a soliton to be well approximated by a solution of 
the linearization of the field equation about $\v$, namely
\beq\label{maw}
\bphi(r,\theta)=\v-\frac{q_1\mu_1}{2\pi}K_1(\mu_1 r)\cos\theta\eps_1
-\frac{q_2\mu_2}{2\pi}K_1(\mu_2 r)\cos\theta\eps_2+\cdots
\eeq
where $K_\nu$ denotes the modified Bessel function of the second kind and 
$q_1,q_2$ are unknown
real constants which will receive a physical interpretation shortly. 
We define linearized fields $\chi_1,\chi_2$ so that 
$\bphi=\v+\chi_1\eps_1+\chi_2\eps_2$
and observe \cite{PSZ} that (\ref{maw}) corresponds to the solution 
of the linearized model, with Lagrangian density
\beq\label{mawifs}
{\cal L}=\sum_{i=1}^2
\bigg(
\frac12\cd_\mu\chi_i\cd^\mu\chi_i-\frac12\mu_i^2\chi_i^2+\kappa_i\chi_i
\bigg)
\eeq
in the presence of external point sources
\beq
\kappa_1=q_1\cd_x\delta(\bx),\qquad
\kappa_2=q_2\cd_y\delta(\bx).
\eeq
Asymptotically, the soliton coincides with the fields induced by 
scalar dipoles of moment
$q_1$ pointing along the $x$-axis and $q_2$ along the $y$-axis, 
inducing fields of mass $\mu_1$ and
$\mu_2$ respectively. 
This is the soliton in standard orientation and position. If the
soliton is rotated through an angle $\alpha$ and translated to 
$\by\in\R^2$, it corresponds to the
composite point source
\beq
\kappa_1=\bq_1\cdot\nabla\delta(\bx-\by),\qquad
\kappa_2=\bq_2\cdot\nabla\delta(\bx-\by),
\eeq
where $\bq_1=q_1(\cos\alpha,\sin\alpha)$ and 
$\bq_2=q_2(-\sin\alpha,\cos\alpha)$. 
The interaction
energy experienced by two solitons placed at $\by$ and 
$\wh{\by}$ with orientations 
$\alpha,\wh\alpha$
is expected to coincide asymptotically as $R=|\by-\wh\by|\ra\infty$
with that of their corresponding point sources $\kappa_i,\wh\kappa_i$
interacting via the linear theory (\ref{mawifs}),
\beq
U=-\int
(\kappa_1\wh{\chi}_1+\kappa_2\wh{\chi}_2)
\,d^2x,
\eeq
where $\wh{\chi}_i$ denotes the field induced by $\wh\kappa_i$. 
A lengthy but straightforward calculation yields the formula
\beq\label{mawifsgf}
U=\frac{1}{2\pi}\sum_{i=1}^2\mu_i^2\left\{K_0(\mu_iR)q_i^\parallel\wh{q}_i^\parallel
+\frac{K_1(\mu_iR)}{\mu_iR}\left[q_i^\parallel\wh{q}_i^\parallel-q_i^\perp\wh{q}_i^\perp\right]\right\},
\eeq
where $\parallel$ and $\perp$ represent the components of 
$\bq$ relative to the orthonormal basis $\bn=(\by-\wh\by)/R$ and $\bn^\perp.$ 

The form of $U$ depends strongly on whether $\mu_1$ and $\mu_2$
are equal.
If $\mu_1<\mu_2$, the expression (\ref{mawifsgf}) is dominated at 
large $R$ by its first term
\beq\label{mawifsgffbrh}
U=\frac{\mu_1^2}{2\pi}K_0(\mu_1R)q_1^\parallel\wh{q}_1^\parallel+\cdots
\eeq
which predicts that the force between two solitons 
is maximally attractive when the 
dominant dipoles
$\bq_1$, $\wh\bq_1$ are anti-aligned along the line joining the 
soliton centres ($\bq_1=-\wh\bq_1=\pm q_1\bn$). 
The neglected terms in (\ref{mawifsgf})
may become significant at intermediate range, 
particularly if $\mu_2^2-\mu_1^2$ is small. 

If $\mu_1=\mu_2$ and $q_1=q_2$, then
whatever the orientations of the two solitons,
 $q_1^\parallel=q_2^\perp$, $q_1^\perp=-q_2^\parallel$ (and
similarly for $\wh\bq_i$), so (\ref{mawifsgf}) simplifies to
\beq\label{mawifsgffb}
U=\frac{\mu_1^2}{2\pi}K_0(\mu_1R)\bq_1\cdot\wh\bq_1,
\eeq
as found in \cite{PSZ}. This is maximally attractive when $\bq_1=-\wh\bq_1$,
independent of the orientation of $\bq_1$ relative to $\bn$. Depending on
the details of $V$, there may be higher order terms which break this symmetry.
 Note that the choice of eigenvectors $\eps_1,\eps_2$ is purely arbitrary 
in this case (since $\mu_1^{-2}J_\v$ is the identity map),
so the notion of standard orientation is similarly a matter of convention.

Given the above, it is interesting to determine what properties of $V,\v$ 
will enforce $\mu_1=\mu_2$.
Let $G\subset O(3)$ denote the subgroup of the isometry group of $S^2$ 
leaving $V$ invariant, and $G_\v\subset G$ the isotropy subgroup of $\v$ in 
$G$. There is an induced isometric action of $G_\v$ on
$T_\v S^2$ which commutes with $J_\v$, so the eigenspaces of $J_\v$ are 
invariant under $G_\v$. Hence,
if $\mu_1\neq\mu_2$, $T_\v S^2$ must have a line 
(in fact, an orthogonal pair of lines) invariant
under $G_\v$. Now $G_\v\subset O(2)$ (the isometry group of $T_\v S^2$), 
and an element of $O(2)$ fixes
a line if and only if it has order 2.\, Hence, unless $G_\v=1$ 
or $\Z_2$ or $\Z_2\times \Z_2$,
$T_\v S^2$ has no such fixed line, and we conclude that $\mu_1=\mu_2$. 
It does not immediately
follow that $U$ simplifies to (\ref{mawifsgffb}) however, 
since this also requires $q_1=q_2$. 
The last condition follows if we assume (as is certainly plausible) 
that the single soliton
is $G_\v$ equivariant, as we will now demonstrate. 

So assume $G_\v$ is nontrivial and different from $\Z_2$ and $\Z_2\times \Z_2$.
 Having chosen eigenvectors $\eps_1,\eps_2$ we have an induced isomorphism 
between the vector spaces $\R^2$ (physical space) and $T_\v S^2$, 
which we can use to transfer the $G_\v$ action to $\R^2$. 
Then a map $\bphi:\R^2\ra S^2$ is $G_\v$ equivariant if
$g\bphi(\bx)=\bphi(g\bx)$ for all $g\in G_\v$. 
Let $\X:\R^2\ra T_\v S^2$ be the asymptotic field 
defined in 
(\ref{maw}) so that $\bphi=\v+\X+\cdots$, namely (since $\mu_1=\mu_2$),
\beq
\X(\bx)=\frac{\mu_1^2}{2\pi}\frac{K_0'(\mu_1r)}{\mu_1r}(q_1x\eps_1+q_2y\eps_2).
\eeq
Equivariance of $\bphi$ implies $g\X(\bx)=\X(g\bx)$, and hence every 
$g\in G_\v$ commutes with
${\rm diag}(q_1,q_2)$. But, if $q_1\neq q_2$, this implies $G_\v$ 
has only elements of order 2,
which, by assumption, is false. 
Hence {\em if the single soliton is $G_\v$ equivariant}, $q_1=q_2$.

The potential studied in the current paper, (\ref{pot2}) with $N\geq 3$,
nicely illustrates this symmetry analysis. If we choose
boundary value $\v=(0,0,1)$ then $G_\v=D_N$ and $\mu_1=\mu_2=m$. 
Further, (at least for $N=3$) 
the single soliton is
observed to be $D_N$ equivariant, so $q_1=q_2$ and the long range forces 
are as described by (\ref{mawifsgffb}). By contrast, if we choose any of 
the other vacua, for example $\v=(1,0,0)$, then
$G_\v=\Z_2$ and there is no reason why $\mu_1$ and $\mu_2$ should be equal. 
The eigenvectors must be invariant under the $\Z_2$ action, which fixes them, 
up to orientation, as $(0,1,0)$ and $(0,0,1)$.
The corresponding eigenvalues are $2N^2m^2$ and $0$, so with our choice of 
conventions, $\mu_1=0$, $\mu_2=\sqrt{2}Nm$, $\eps_1=(0,0,1)$, 
$\eps_2=(0,-1,0)$, and the long range forces are as described in \cite{JS}. 

It would be interesting to see how this analysis generalizes to the 
$(3+1)$-dimensional Skyrme model.

\section*{Appendix B: the stress-energy of a baby Skyrme lattice}

In sections \ref{sollat} and \ref{sec-alt}, baby Skyrme models with 
doubly periodic boundary conditions were studied. Equivalently, the model
was put on a torus $\C/\Lambda$, where $\Lambda$ is a period lattice, chosen
in this case to be rhombic, 
$\Lambda_*=\{(n+me^{i\pi/3})L\: :\: n,m\in\Z\}$, where the side length $L>0$
is determined numerically. Of course, given {\em any}
lattice $\Lambda$, and any topological charge $B$, one would
expect the baby Skyrme model on $\C/\Lambda$ to have an energy minimizer
of charge $B$. Not all such doubly periodic solutions can be meaningfully
interpreted as soliton lattices, however. For fixed $B$, we have a map
which sends the lattice $\Lambda$ to the energy of its charge $B$ minimizer,
and to be a genuine soliton lattice, $\Lambda$ should be (at least a local)
minimizer of this map. That is, the energy of a soliton lattice should
be stationary under variations not just of the field $\phi$, but also of
the lattice $\Lambda$. This condition can be usefully reformulated
in terms of the stress-energy of the field $\phi$, as we now show.

All tori $\C/\Lambda$ are diffeomorphic via real linear maps, 
so we can fix a standard lattice,
for example $\Lambda_*$, and consider every other torus $\C/\Lambda$ to
be identified with $\C/\Lambda_*$, but with a nonstandard metric $g$ (the
pullback of the usual metric on $\C/\Lambda$ by the diffeomorphism
$\C/\Lambda_*\ra\C/\Lambda$). So now the domain of $\phi$, call it
$M,$ is fixed as a smooth manifold, but has a Riemannian metric $g$ which
varies as we vary $\Lambda$. In order to be a soliton lattice, 
a field $\phi:(M,g)\ra S^2$ should be a critical point of $E$ under 
all smooth variations
of $\phi$, and all variations of $g$ arising from changing $\Lambda$.
This leads us to compute the variation of $E(\phi,g)$ with respect to $g$.

It costs no effort to put the computation in a general geometric setting. So
let $M$ be a compact oriented $n$-manifold, $(N,h)$ be a Riemannian manifold,
$\omega$ be an $n$-form on $N$ and $V:N\ra\R$ be a smooth function. For a given
metric $g$ on $M$, the energy of a map $\phi:M\ra N$ is, by definition,
\beq
E(\phi,g)=\frac12\|\d\phi\|^2+\frac12\|\phi^*\omega\|^2+\int_MV(\phi)\,\vol_g
\eeq
where $\|\cdot\|$ denotes $L^2$ norm and $\vol_g$ is the volume form on $M$
associated with $g$. We wish to compute the variation of $E(\phi,g)$ with 
respect to $g$. Let $g_t$ be a smooth curve in the space of Riemannian
metrics, with $g_0=g$ and $\cd_tg_t|_{t=0}=\varepsilon$. Note that
$\varepsilon$ is (like $g$) a section of $\odot^2T^*M$, a real vector
bundle over $M$ of rank $\frac12n(n+1)$. This bundle inherits a fibre metric
from $g$, which we denote $\ip{\cdot,\cdot}$, 
defined as follows: let $e_1,\ldots,e_n$ be a local orthonormal 
coframe on $(M,g)$; then we demand that 
$\{\frac12(e_i\otimes e_j+e_j\otimes e_i)\: :\: i\leq j\}$
is a local orthonormal frame for $(\odot^2T^*M,\ip{\cdot,\cdot})$. 
The key fact is:

\vspace*{0.3cm}
\noindent
{\bf Proposition}\,
 Let $g_t$ be a smooth, one-parameter family of metrics on $M$
with $g_0=g$ and $\cd_tg_t|_{t=0}=\varepsilon$. Then, for fixed
$\phi:M\ra N$,
$$
\left.\frac{d\: }{dt}E(\phi,g_t)\right|_{t=0}=\frac12\ip{S(\phi),\varepsilon}_{L^2}
$$
where
$$
S(\phi)=(\frac12|\d\phi|^2
-\frac12|\phi^*\omega|^2+V(\phi))g-\phi^*h
$$
which, following the terminology of harmonic
map theory, we call the {\em stress-energy tensor}. Note that, like $g$ and
$\varepsilon$, $S(\phi)$ is a section of $\odot^2T^*M$.

\vspace*{0.3cm}
\noindent
{\it Proof:} We compute separately
the variations of the three terms in $E$, which we denote $E_2,E_4$ and $E_0$
respectively.
The first term $E_2=\frac12\|\d\phi\|^2$ is the Dirichlet energy of $\phi$,
whose variation with respect to $g$ is \cite{baiwoo}
\beq
\left.\frac{d\: }{dt}E_2(\phi,g_t)\right|_{t=0}=\frac12\ip{\frac12|\d\phi|^2g
-\phi^*h,\varepsilon}_{L^2}.
\eeq
In the course of the proof of this, one finds that
\beq\label{B1}
\cd_t|_{t=0}\vol_{g_t}=\frac12\tr\varepsilon\, \vol_g=
\frac12\ip{\varepsilon,g}\vol_g.
\eeq
It follows that the third term, $E_0=\int_MV(\phi)\vol_g$, has variation
\beq
\left.\frac{d\: }{dt}E_0(\phi,g_t)\right|_{t=0}=\frac12\ip{V(\phi)g,\varepsilon}_{L^2}.
\eeq
Finally, to handle the middle term $E_4$,
we must compute the variation of the Hodge isomorphism
$*_t:\Omega^n(M)\ra\Omega^0(M)$ defined by $g_t$. Let $\mu$ be a fixed $n$-form 
on $M$. Then, by definition $*_t\mu=f_t$ where $\mu=f_t\vol_{g_t}$. 
Differentiating this with respect to $t$ at $t=0$ yields
\beq
0=(\cd_t f_t)|_{t=0}\vol_g+\frac12f_0\ip{\varepsilon,g}\vol_g
\eeq
using (\ref{B1}). Hence
\beq
\cd_t(*_t\mu)|_{t=0}=-\frac12\ip{\varepsilon,g}*_0\mu
\eeq
and so
\beq
\left.\frac{d\: }{dt}E_4(\phi,g_t)\right|_{t=0}=
\frac12\frac{d\: }{dt}\int_M\phi^*\omega\wedge *_t\phi^*\omega
=-\frac14\int_M\ip{\varepsilon,g}\phi^*\omega\wedge *\phi^*\omega
=-\frac14\ip{\varepsilon,|\phi^*\omega|^2g}_{L^2}
\eeq
which completes the proof.
\vspace*{0.3cm}

\begin{figure}[ht]\begin{center}
\includegraphics[width=6cm]{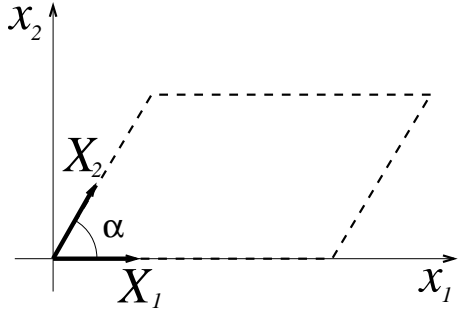}
\caption{The period parallelogram of a generic torus.}
\label{fig-cl}\end{center}\end{figure}

We now return to the problem of finding criticality constraints on a 
baby Skyrmion lattice. So from now on, $M=\bb{T}^2=\C/\Lambda$, 
where $\Lambda$ is some fixed lattice, $N=S^2$, 
$\omega=\kappa\times(\mbox{area form on $S^2$})$ and
$V$ is some potential. Without loss of generality, we can
assume that one of the periods of $\Lambda$ is positive-real, and the other
has argument $\alpha\in(0,\pi)$. Then the period parallelogram is as
depicted in Figure~\ref{fig-cl}.
There is a three-parameter family of 
variations of $\Lambda$, up to isometries, generated by (a) homothety
(uniformly scaling the parallelogram), 
(b) stretching the parallelogram horizontally, and (c)
varying the interior angle of the parallelogram through $\alpha$. 
We claim that the 
corresponding 
variations of the metric $g=dx_1^2+dx_2^2$ are tangent to the symmetric
bilinear forms
\bea
\varepsilon_{(a)}&=&g\nonumber\\
\varepsilon_{(b)}&=&dx_1^2\nonumber\\
\varepsilon_{(c)}&=&-2dx_1dx_2+\cot\alpha dx_2^2.\label{B2}
\eea
Of these, $\varepsilon_{(a)}$ and $\varepsilon_{(b)}$ are clear. To obtain
$\varepsilon_{(c)}$, we note that varying $\alpha$ as $\alpha+t$ is equivalent
to defining the inner product between the fixed pair of unit vectors
$X_1,X_2$ to be $\cos(\alpha+t)$, while keeping their lengths unchanged
(see Figure~\ref{fig-cl}). 
Hence,
\beq
\varepsilon(X_1,X_1)=\varepsilon(X_2,X_2)=0,\qquad
\varepsilon(X_1,X_2)=\varepsilon(X_2,X_1)=-\sin\alpha.
\eeq
Now 
\beq
\frac{\cd\: }{\cd x_1}=X_1\qquad\mbox{and}\qquad
\frac{\cd\: }{\cd x_2}=\frac{1}{\sin\alpha}(X_2-\cos\alpha X_1),
\eeq
so 
\beq
\varepsilon(\cd/\cd x_1,\cd/\cd x_1)=0,\quad
\varepsilon(\cd/\cd x_1,\cd/\cd x_2)=-1,\quad
\varepsilon(\cd/\cd x_2,\cd/\cd x_2)=\cot\alpha
\eeq
which is equivalent to (\ref{B2}).
If $\phi:M\ra S^2$ is a soliton lattice, then, by the
Proposition, $S(\phi)$ must be $L^2$ orthogonal
to each of $\varepsilon_{(a)},\varepsilon_{(b)},\varepsilon_{(c)}$.
Hence, $S(\phi)$ must be $L^2$ orthogonal to any section of $\odot^2T^*M$
in the span of these, for example
\beq
\varepsilon_{(a)}=g,\qquad
\varepsilon_{(b')}=dx_1dx_2,\qquad
\varepsilon_{(c')}=dx_1^2-dx_2^2.
\eeq
 Now $\ip{g,g}=n=2$, and $\ip{\phi^*h,g}=|\d\phi|^2$, so
\beq
\ip{S(\phi),\varepsilon_{(a)}}_{L^2}=0\quad\Leftrightarrow\quad
\int_M\left(-\frac12|\phi^*\omega|^2+V(\phi)\right)\vol_g=0,
\label{B3}\eeq
which coincides with identity (\ref{identity1}). Further, 
$\ip{\varepsilon_{(b')},g}=\ip{\varepsilon_{(c')},g}=0$, so
\beq
\ip{S(\phi),\varepsilon_{(b')}}=0\quad\Leftrightarrow\quad
\ip{\phi^*h,dx_1dx_2}=\int_M\frac{\cd\phi}{\cd x_1}\cdot\frac{\cd\phi}{\cd x_2}
dx_1\, dx_2=0
\label{B4}\eeq
which is identity (\ref{identity3}), and
\beq
\ip{S(\phi),\varepsilon_{(c')}}=0\quad\Leftrightarrow\quad
\ip{\phi^*h,dx_1^2-dx_2^2}=
\int_M\left(\left|\frac{\cd\phi}{\cd x_1}\right|^2
-\left|\frac{\cd\phi}{\cd x_2}\right|^2\right)
dx_1\, dx_2=0
\label{B5}\eeq
which is identity (\ref{identity2}).  

We conclude by making two remarks. First, it is interesting that we
get the {\em same} integral constraints on a baby Skyrmion lattice for
{\em all} tori $\C/\Lambda$. Second, in the case where the minimizer
$\phi$ is holomorphic (e.g.\ the alternative $D_3$ model in section
\ref{sec-alt}, or Ward's model \cite{Wa}) only the scaling constraint
(\ref{B3}) is nontrivial, since $\phi$ is conformal and
$\cd/\cd x_1$, $\cd/\cd x_2$ are orthonormal, so $\cd \phi/\cd x_1$ and
$\cd\phi/\cd x_2$ have equal length and are orthogonal, pointwise, and hence
(\ref{B4}) and (\ref{B5}) hold automatically.

\section*{Note added}
The preprint \cite{Ma} contains several ideas related to
those appearing in this paper. In particular, there is
a similar proposal to identify quarks inside Skyrmions,
and a suggestion that many new Skyrmions might be found
as pieces of the Skyrme crystal; in the same way that
the polyiamond solitons may be viewed as pieces of the
soliton lattice.

\section*{Acknowledgements}
\noindent 
Many thanks to Richard Ward for useful discussions.
We acknowledge EPSRC and STFC for grant support.

\end{document}